# ImmunoNX: a robust bioinformatics workflow to support personalized neoantigen vaccine trials

**Authors**
Kartik Singhal[*,1], Evelyn Schmidt[*,1], Susanna Kiwala[1], S. Peter Goedegebuure[2,4], Christopher A. Miller[1,2], Huiming Xia[1], Kelsy C. Cotto[1], Jinglun Li[1], Jennie Yao[1], Luke Hendrickson[1], Miller M. Richters[1], My H. Hoang[1], Mariam Khanfar[1], Isabel Risch[1], Shelly O'Laughlin[3], Nancy Myers[2,4], Tammi Vickery[8], Sherri R. Davies[4], Feiyu Du[3], Thomas B. Mooney[3], Adam Coffman[1], Gue Su Chang[3], Jasreet Hundal[3], John E. Garza[3], Michael D. McLellan[3], Joshua F. McMichael[1], John Maruska[3], William Blake Inabinett[6], William A. Hoos[6], Rachel Karchin[7], Tanner M. Johanns[1,2], Gavin P. Dunn, Russel K. Pachynski[1,2], Todd A. Fehniger,[1] Jeffrey P. Ward[1,2], Jennifer A. Foltz[1,2], William E. Gillanders[†,2,4], Obi L. Griffith[†,1,2,3,5], Malachi Griffith[†,1,2,3,5]

**Affiliations**
[1]Division of Oncology, Department of Medicine, Washington University School of Medicine, St. Louis, MO, USA
[2]Siteman Cancer Center, Washington University School of Medicine, St. Louis, MO, USA
[3]McDonnell Genome Institute, Washington University School of Medicine, St. Louis, MO, USA
[4]Department of Surgery, Washington University School of Medicine, Saint Louis, MO, USA
[5]Department of Genetics, Washington University School of Medicine, St. Louis, MO, USA
[6]Jaime Leandro Foundation for Therapeutic Cancer Vaccines, Chapel Hill, NC, USA
[7]Johns Hopkins Department of Biomedical Engineering, Baltimore, MD, United States
[8]Department of Pathology and Immunology, Washington University School of Medicine, St Louis, MO, USA

## Abstract

Personalized neoantigen vaccines represent a promising immunotherapy approach that harnesses tumor-specific antigens to stimulate anti-tumor immune responses. However, the design of these vaccines requires sophisticated computational workflows to predict and prioritize neoantigen candidates from patient sequencing data, coupled with rigorous review to ensure candidate quality. While numerous computational tools exist for neoantigen prediction, to our knowledge, there are no established protocols detailing the complete process from raw sequencing data through systematic candidate selection. Here, we present ImmunoNX (Immunogenomics Neoantigen eXplorer), an end-to-end protocol for neoantigen prediction and vaccine design that has supported over 185 patients across 11 clinical trials. The workflow integrates tumor DNA/RNA and matched normal DNA sequencing data through a computational pipeline built with Workflow Definition Language (WDL) and executed via Cromwell on Google Cloud Platform. ImmunoNX employs consensus-based variant calling, in-silico HLA typing, and pVACtools for neoantigen prediction. Additionally, we describe a two-stage immunogenomics review process with prioritization of neoantigen candidates, enabled by pVACview, followed by manual assessment of variants using the Integrative Genomics Viewer (IGV). This workflow enables vaccine design in under three months. We demonstrate the protocol using the HCC1395 breast cancer cell line dataset, identifying 78 high-confidence neoantigen candidates from 322 initial predictions. Although demonstrated here for vaccine development, this workflow can be adapted for diverse neoantigen therapies and experiments. Therefore, this protocol provides the research community with a reproducible, version-controlled framework for designing personalized neoantigen vaccines, supported by detailed documentation, example datasets, and open-source code.

# Introduction

Personalized neoantigen vaccines utilize a combination of immunotherapy and genomics to combat cancer. A subset of tumor-specific somatic alterations generate neoantigens that bind to MHC molecules and are presented directly on the tumor cell's surface and by professional antigen-presenting cells (APCs) for recognition by T cells[1]. T-cell based immunotherapies leverage this fact to use the patient's T cells to kill the patient's tumor. *In silico* techniques can be used to predict the neoantigens most amenable to targeting, and these can be packaged into a vaccine that, when administered to the patient, trains their immune system to identify and kill tumor cells. Currently, there are over 100 interventional trials on clinicaltrials.gov involving cancer vaccines. The Food and Drug Administration (FDA) granted a breakthrough therapy designation to the combination treatment of Pembrolizumab and V940, a personalized mRNA vaccine[2], for high-risk melanoma, with phase III trials underway.

      We have previously developed bioinformatics tools for neoantigen analysis including the pVACtools suite ([pvactools.org](pvactools.org))[3–7], and described best practices and recommendations for neoantigen selection[8,9]. At our institute alone, pVACtools has been used to design neoantigen vaccines for at least 185 patients involved in 11 completed or ongoing clinical trials (NCT03606967, NCT03199040, NCT05111353, NCT03532217, NCT04397003, NCT03122106, NCT04015700, NCT03956056, NCT05743595), published clinical trials (NCT03121677[10], NCT02348320)[11], and case studies[12–15]. pVACtools has demonstrated remarkable community adoption, with over 164,000 downloads from Python Package Index (PyPI) and over 113,000 Docker pulls from DockerHub. This widespread usage has facilitated studies of immune evasion[16,17], tumor microenvironment evolution[18], the relationship between neoantigen burden and prognosis[19–28], mechanisms of response to immune checkpoint blockade therapy[29–31], development of mRNA vaccines[32],[33–36] characterization of the neoantigen landscape across numerous tumor types[37–40], algorithm development, and many more. Despite this widespread adoption of neoantigen vaccine-related treatments and computational tools, there remains a lack of established and comprehensive computational protocols that address the complexities associated with the design of personalized cancer vaccines. These complexities include integrating predictions from multiple binding algorithms that often disagree, ensuring mutant peptide sequences don't inadvertently match wildtype sequences from other human proteins, maintaining version control to allow continuous pipeline improvements while preserving consistency within clinical trials, providing a user-friendly interface to evaluate neoantigens, and incorporating manual review to validate computational predictions. Here, we present a detailed procedure using the computational pipeline ImmunoNX (Immunogenomics Neoantigen eXplorer), which already supports clinical trials at Washington University in St Louis; and make it available to facilitate the design of personalized neoantigen-based immunotherapy trials initiated by the research community. While this protocol focuses on neoantigen vaccine design, ImmunoNX is applicable to additional therapeutic applications including personalized adoptive T cell therapies, engineered TCR T cell therapies, and the creation of immunogenicity validation datasets for research purposes.

## Overview of the protocol

We have leveraged our previously developed neoantigen prediction tools[5], genomics analysis pipelines[41], and experience from ongoing/completed clinical trials to develop a protocol for designing neoantigen vaccines. The workflow starts with the patient's tumor DNA/RNA and matched-normal DNA sequencing data and ends with a neoantigen vaccine, ready for manufacture (**Figure 1**). Our protocol can broadly be divided into two steps: (1) ImmunoNX, a comprehensive computational immunogenomics pipeline for neoantigen prediction, and (2) a rigorous prioritization and immunogenomics review of predicted candidates. The protocol relies on several independently validated tools (**Supplementary Table 1**) and publicly available reference sequence and annotation files. The immunogenomics review step provides critical checks of the predicted neoantigen candidates to ensure safety and the highest likelihood of therapeutic success. ImmunoNX is continuously improved by the addition of new tools and features. Prior to releasing updates, the pipeline is run on multiple datasets including the well-characterized tumor/normal cell-line (HCC1395), presented here, and real-world patient data. Results from these test runs are then compared against the previous version's results using the pVACcompare tool (available as part of the pvactools Python package version 6.0. and above).

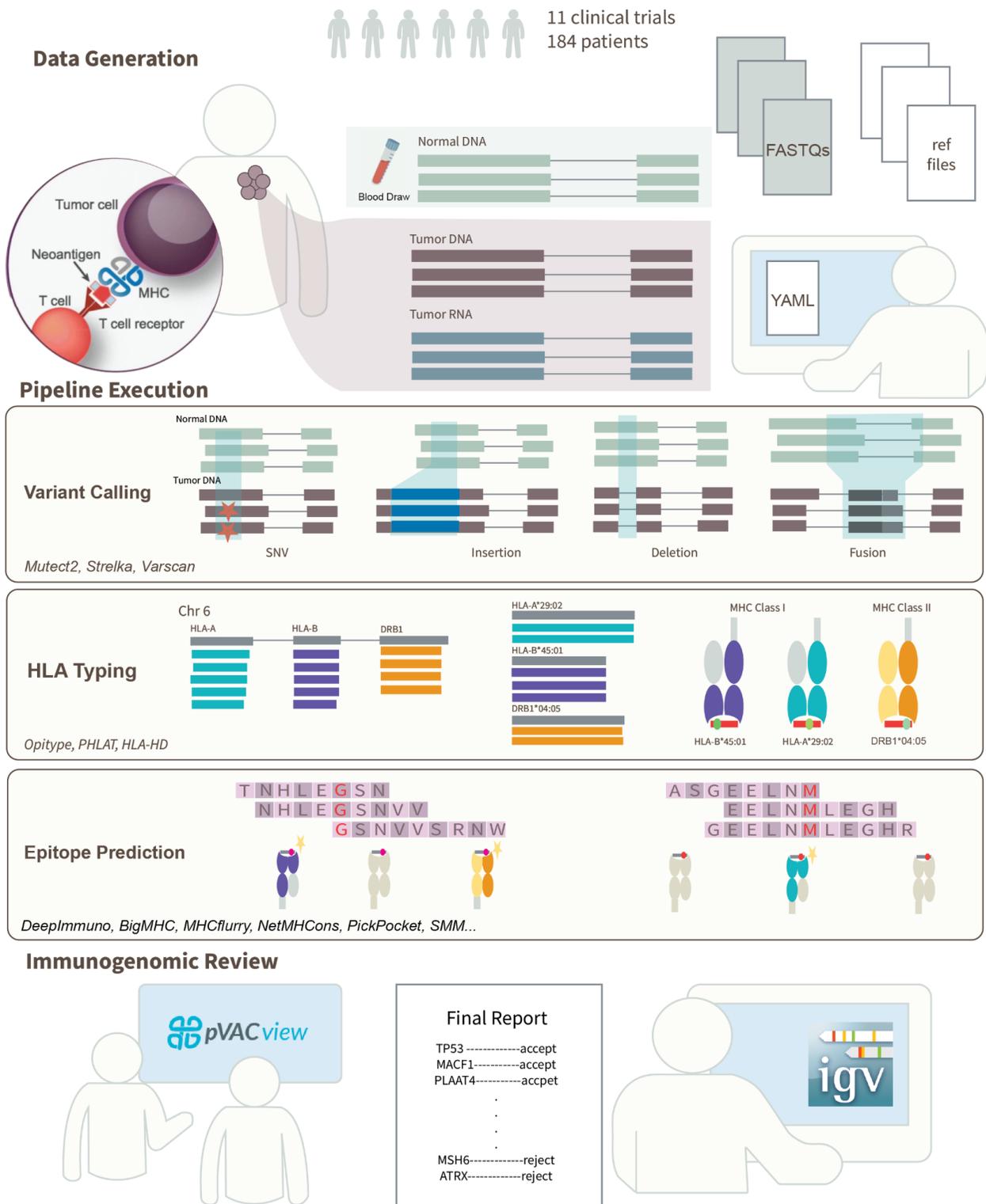

**Figure 1: Overview of immunogenomics pipeline for neoantigen prediction and personalized cancer vaccine development.** This workflow has been refined over 11 clinical trials. It begins with 'normal' (usually blood) and tumor sample collection. Data generation includes sequencing of normal DNA, tumor DNA, and tumor RNA. The pipeline execution phase involves variant calling to identify somatic single nucleotide variants, and small insertions and deletions. In-silico HLA typing is conducted to determine the patient's HLA alleles. Various epitope prediction algorithms predict MHC Class I and Class II binding affinity, presentation, and immunogenicity scores for peptides

generated from the identified variants. The immunogenomics review phase utilizes pVACview for prioritization of candidates, and the Integrated Genomics Viewer (IGV) to ensure the selected candidates are truly expressed at the DNA and RNA levels. Finally, the reviewer generates a report that can be shared with the appropriate vaccine manufacturer, and the vaccine can be administered to the patient.

ImmunoNX uses the Workflow Definition Language (WDL) to define workflow steps, resource requirements for each task, and dependencies between tasks. The pipeline takes advantage of Docker images to be highly portable, and therefore can be run on a local workstation, a local High Performance Cluster (tested on LSF and Slurm), or on a cloud platform (tested on Google Batch and Terra). These WDL workflows and helper scripts are publicly available on GitHub (and archived on Zenodo) and are executed using the cromwell workflow management system[42]. ImmunoNX requires three sets of sequencing data for each patient: DNA from (1) tumor and (2) healthy tissue (with support for whole exome, whole genome, or targeted DNA sequencing), and (3) RNA from the tumor tissue. We provide our sequencing depth recommendations for each in **Supplementary Table 2**. If clinical HLA typing was performed on the patient, that may be used as an additional input to the pipeline, however, the pipeline also carries out in silico HLA class I and class II predictions. While the pipeline can be run on various platforms, in this manuscript, we demonstrate our public and open (MIT) license Google Cloud implementation of the pipeline (**Figure 2**).

Once a Google Cloud account is set up, the workflow begins by downloading GitHub repositories with Google Cloud helper scripts. These scripts enable the upload of sequencing data (raw FASTQ files or unaligned BAMs) to a Google bucket and set up of a virtual machine (VM) from which the pipeline is launched. An initial data quality control check is performed to ensure adequate quality of the sequencing data, HLA allele concordance between in-silico prediction tools and clinical data (if available), sample identity verification, and sufficient tumor purity. The pipeline outputs include DNA and RNA alignments, class I and II HLA genotypes, germline and somatic variant calls, transcript expression estimates, and predicted neoantigen candidates. In addition to gene and transcript expression information, the variants and neoantigen candidates files are also annotated with various databases including dbSNP[43], ClinVar[44], and gnomAD[45]. The neoantigen candidates are then reviewed using the pVACview[7] application and prioritized to select up to a target number of neoantigen candidates with the highest likelihood of stimulating an immune response. A subsequent immunogenomics review phase involves rigorous examination of: sequence data support for the variant giving rise to the neoantigen at the DNA and RNA levels using the Integrative Genomics Viewer (IGV)[46] and MHC binding, presentation, and immunogenicity prediction algorithms. A genomics review report summarizes all findings, quality metrics and special considerations for individual candidates (see **Supplementary File 1** for example report). Finally, a vaccine order form generated by pVACtools is reviewed, and can be sent to the appropriate vaccine manufacturer.

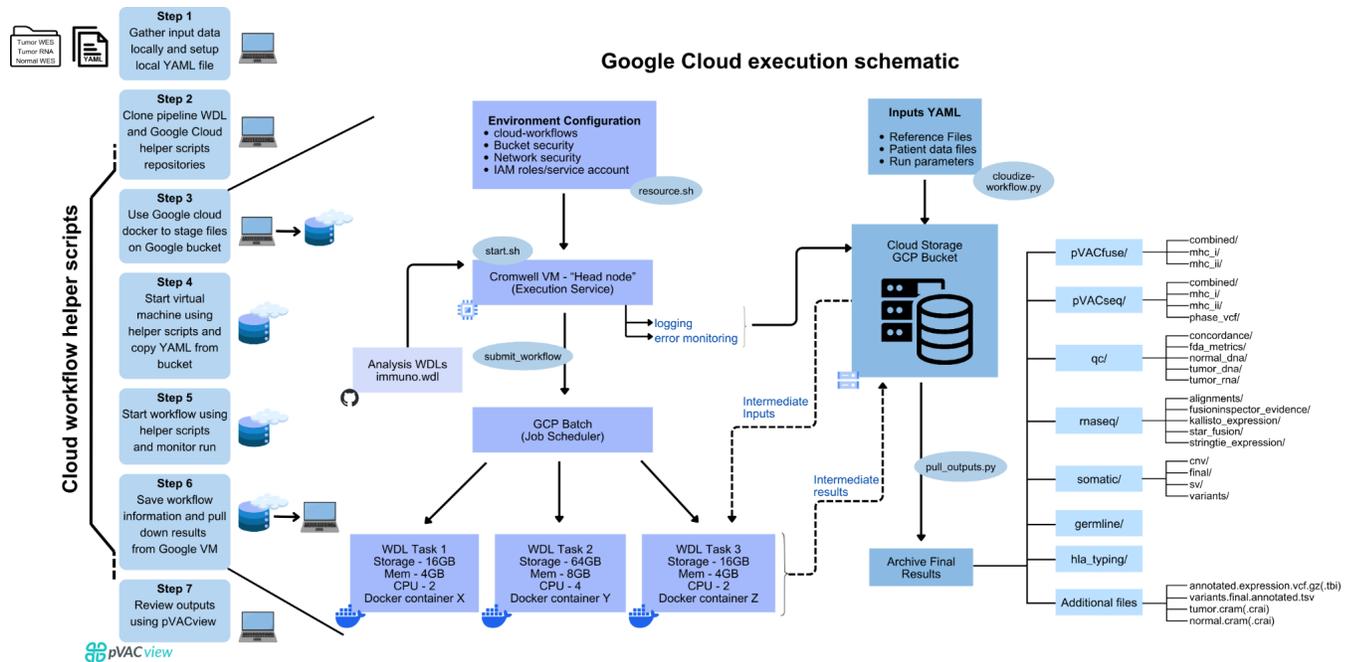

**Figure 2: Google Cloud implementation for the ImmunoNX pipeline**. The workflow comprises seven key steps, from initial data gathering to generating files for final review. Step 1 involves the local data preparation and YAML file set-up on a laptop or local server. In step 2, the Cloud workflow helper scripts are downloaded, which will allow the user to interact with the Google Cloud virtual machine and storage bucket. Steps 3 and 4 use these scripts to configure the Cloud environment, transfer input data files to the Cloud bucket, and set up the head node virtual machine (VM) that runs the workflow management system, Cromwell. The 'resources.sh' script configures the environment, and creates the billing project and bucket, if they do not already exist. The 'start.sh' script creates the VM and copies the analysis-wdls repository which contains WDL files required for running the workflow and some additional helper scripts to the VM. In step 5, the 'submit_workflow' command from the helper script is used to initiate the workflow, wherein Cromwell parses the WDL files and submits jobs to the job scheduler, GCP Batch. The VM actively generates a log file for the entire run that is copied to the Cloud bucket at the end of the run (regardless of whether the run fails or succeeds). Additionally, intermediate files generated from the tasks are continuously being transferred to the bucket and being read from the bucket. In step 6, on successful completion, workflow information is saved, and the final results are retrieved from the Cloud bucket using the 'pull_outputs.py' script. Finally, in step 7 the neoantigen predictions output from the pipeline can be visualized interactively using pVACview.

## Neoantigen prediction implementation

Neoantigen prediction begins with the identification of somatic variants through a tumor-normal variant calling pipeline. This pipeline aligns raw DNA FASTQ files and integrates predictions from multiple somatic variant callers, including Mutect2[47], Strelka[48], and Varscan[49], and annotates known cancer variants with DoCM[50], to enable high-confidence consensus calling of missense, in-frame and frameshift indels, and di-nucleotide variants (**Figure 3, Supplementary Figure 1**). This list of variants is further filtered based on various criteria such as read depth, mapping quality of reads, and gnomAD frequency of the variant. Simultaneously, the normal DNA FASTQ files are also processed to identify germline variants with GATK[51] in the patient and identify HLA alleles with Optitype[52], PHLAT[53], and HLA-HD[54]. Tumor RNA FASTQ files are aligned, and gene and transcript isoform expression values are calculated using Kallisto[55] and StringTie[56], along with gene fusion detection using STAR-fusion[57] and annotation using AGfusion[58]. As indicated by the example timeline (**Figure 3**), the ImmunoNX pipeline has been optimized to run the workflows in parallel to ensure efficiency.

Outputs from the above are utilized by pVACtools, a comprehensive computational suite designed to predict and prioritize tumor-specific candidate neoantigens. Cases where germline missense variants are proximal to somatic variants may alter the resulting peptide sequences[4] and are handled appropriately by pVACtools. The RNAseq data is incorporated to allow targeting of neoantigens where the variant-carrying allele is expressed. The neoantigen prediction tools within pVACtools are pVACseq and pVACfuse, which identify neoantigens arising from somatic variants and gene fusions, respectively. pVACtools evaluates all possible peptides arising from somatic variants or gene fusions. It first uses an ensemble of MHC Class I and II binding, elution, and immunogenicity prediction algorithms (**Supplementary Table 1**) to score each potential peptide-HLA pair. Subsequently, it uses these scores along with features such as variant allele expression, transcript expression and support level, anchor residues, and user-provided problematic amino acids to prioritize neoantigen candidates. The pVACview interface allows users to visualize, filter, and select neoantigens for clinical and research applications, making it easy to prioritize candidates of interest. Additional helper scripts and commands allow the generation of a final report with a list of candidates for the specific vaccine design modality.

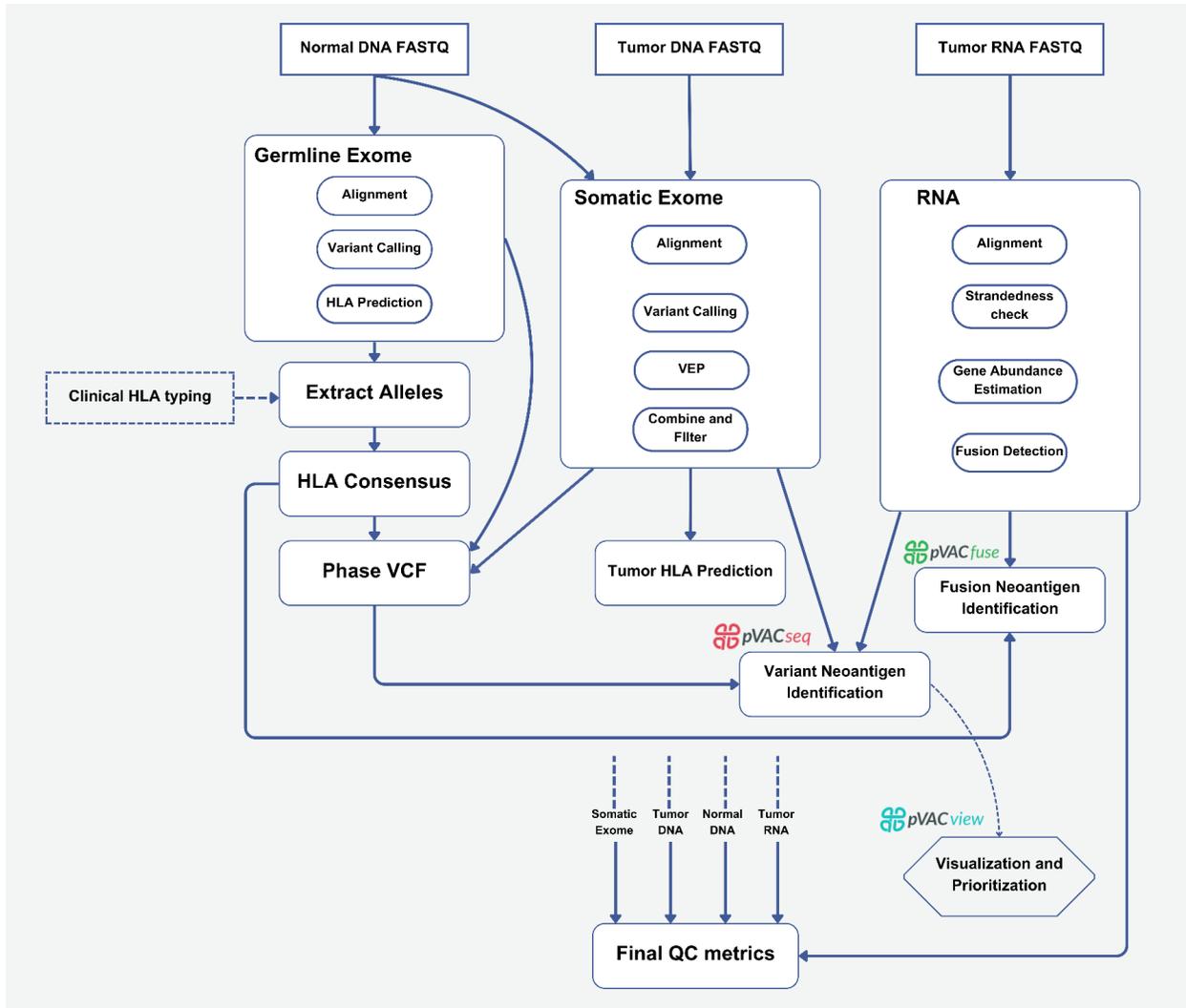

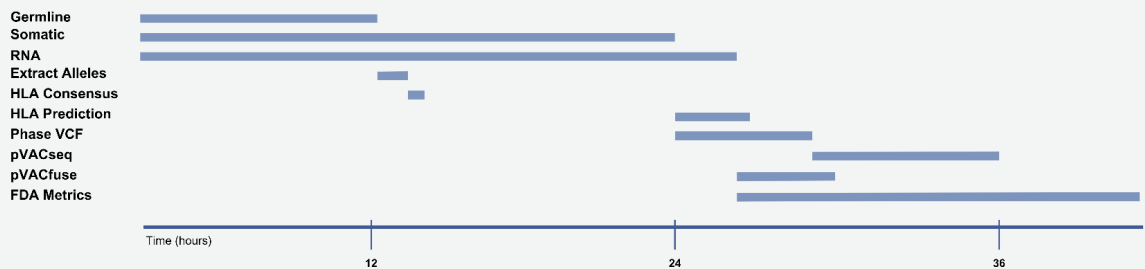

**Figure 3: ImmunoNX pipeline for neoantigen prediction.** The flowchart illustrates the bioinformatics workflow that processes patient sequencing data to predict potential neoantigens. The pipeline takes three primary inputs: tumor DNA, (matched) normal DNA, and tumor RNA sequencing data, with an option to incorporate clinical HLA typing data, if available. The normal and tumor DNA data are aligned and used for somatic variant calling, through the somatic variant calling pipeline, which uses a consensus variant calling approach. The normal DNA data is used for class I and class II HLA typing with OptiType, PHLAT, and HLA-HD. This HLA information is used in the absence of clinical HLA typing results. Additionally, the normal DNA data is used for germline variant calling, which is essential for identifying in-phase germline variants that are proximal to the somatic variant when predicting neoantigens. The tumor RNA sequencing data is aligned, and its gene and transcript expression is calculated. Gene fusion predictions

are also made from the tumor RNA and used by pVACfuse to identify candidates arising from gene fusions. Finally, pVACseq uses the HLA type data, phased VCF, somatic pipeline VCF, and RNA expression data, to predict neoantigen candidates. Importantly, the pipeline also compiles comprehensive quality control metrics throughout the process, which are recorded in the final report. The pipeline also maintains a record of how it was run, making all this information available to the user, including the timing diagram shown below the flowchart. This displays the duration taken by each step and also shows that Cromwell ensures efficient execution of the pipeline by parallelizing steps.

## Comparison with other approaches

Several computational methods have been developed for neoantigen prediction and vaccine design from sequencing data, each contributing valuable approaches to the field, including TruNeo[59], pTuneos[60], NeoDisc[61], CloudNeo[62], LENS[63], NeoPredPipe[64] and OpenVax[65]. These platforms differ in aspects such as the types of input data they accept, the variant types analyzed, whether they incorporate expression data, the number of prediction algorithms used, deployment environment (local vs cloud-based), and the extent of downstream visualization or prioritization tools provided.

The ImmunoNX pipeline described here offers several features designed to maximize prediction accuracy and clinical utility. Its end-to-end workflow design, starting with processing of raw DNA and RNA sequencing data to downstream neoantigen prediction and prioritization. The integration of expression data is not standard across all tools and provides valuable information for candidate prioritization due to factors such as allele-specific expression and the potential for transcript isoforms that may skip the variant carrying exon. Our pipeline integrates multiple established algorithms for variant calling, binding affinity, presentation, and immunogenicity prediction, providing ensemble-based predictions that leverage the strengths of different computational approaches. Additionally, ImmunoNX incorporates MHC class I and II binding predictions allowing users to prioritize candidates capable of generating both CD4 and CD8 T cell responses. The pVACview interface allows for interactive exploration of candidates, facilitating the integration of expert knowledge into the prediction process during immunogenomics review. Moreover, the modular nature of pVACtools allows for continuous improvement and integration of new methods as the field evolves, such as allowing the addition of tools that prioritize based on presentation or immunogenicity as those algorithms become available. Finally, our emphasis on robust software engineering practices sets this pipeline apart. Docker containers and WDL pipelines allow for easy execution, version control allows users to ensure they have a fixed version of the pipeline, and the comprehensive documentation and open-source nature of the project empowers users to quickly adopt and integrate this pipeline into their trials. To our knowledge, this represents the first step-by-step protocol for neoantigen prediction and review, providing a fully reproducible workflow accompanied by detailed documentation, example datasets, and interactive visualization tools.

## Limitations of the protocol

This neoantigen vaccine design workflow has proven effective in multiple clinical trials, however, it has its limitations. One significant constraint is the reliance on manual steps, particularly in the immunogenomics review phase (**Figure 1**). While this step is valuable as it incorporates expert knowledge to ensure inclusion of vaccine candidates with the highest probability of being immunogenic, it is also time consuming. Therefore, in addition to the existing prioritization features in pVACview, a machine learning-based predictor is being developed to further aid in neoantigen classification (manuscript in preparation).

ImmunoNX is not compatible with long-read sequencing technologies, which could improve the detection of structural variants and complex genomic rearrangements that may give rise to neoantigens[66]. Similarly, neoantigens arising from transposable elements[67] or viral antigens[68] are also not captured by the present workflow. The pipeline also currently does not integrate additional orthogonal datasets such as ribosome profiling (Ribo-seq) which can confirm active translation of variant-containing transcripts[69]; or immunopeptidomics data, which can directly identify MHC-bound peptides[70]. Post-translational modifications, which can generate novel epitopes independent of genomic variation[71], are also not considered in the current pipeline. Additionally, while the pipeline currently does not prioritize neoantigens arising from splice variants, a pVACsplice module for predicting splice neoantigens will be added to ImmunoNX in the near future.

Finally, while the neoantigen vaccine design workflow concludes with the identification and prioritization of candidate neoantigens, it does not fully address the vaccine manufacturing stage. In practice, additional communication is required with the appropriate vaccine manufacturers based on the delivery strategy (DNA, peptide, RNA, dendritic cell), to finalize vaccine designs.

## Materials/Equipment

### Input data files

The input files needed by ImmunoNX can broadly be divided into case-specific files and reference/annotation files. While all the files must eventually be uploaded to the Google Cloud bucket, it is recommended that the reference files are stored in a separate read-only bucket. The Preparing_references guide (**Supplementary File 2**) has instructions on compiling reference files using publicly available resources. Alternatively, the reference files available in the 'gs://griffithlab-workflow-inputs' public google bucket can be used by copying the files to the Cloud storage bucket. These reference files are for human genome build 38[72], with a fix enabling detection of variants in the U2AF1 locus[73], and Ensembl version 105[74]. The case-specific files can be kept locally and uploaded to a bucket using the cloudize-workflow script described below and removed after the pipeline has completed and the data has been downloaded. File paths to all the files needed by the pipeline along with configurable run time parameters are summarized in the input YAML file.

The first section of the YAML file outlines the patient-specific data required to run the immunogenomics pipeline. This includes the input files for the tumor DNA sequence, normal DNA sequence, and tumor RNA sequence data. The YAML file also captures metadata about these samples, such as sample names, sequence metadata, and strandedness information for the RNA-seq data. The file optionally allows for the inclusion of any known clinical HLA typing information for the patient. The latter section of the YAML file specifies the various reference files and parameters required for the pipeline. This includes paths to the reference genome, aligner indices, transcript annotations, known variant files, and parameters such as binding thresholds, DNA VAF thresholds, amino acids to avoid in vaccine manufacture and more.

### Scripts/GitHub repositories

ImmunoNX's implementation relies on two GitHub repositories. The analysis-WDLs repository has a collection of workflows for analysis of genomic sequencing data. The cloud-workflows repository contains helper scripts that facilitate the execution of the pipeline on GCP. Additional QC scripts and pipeline helper commands can compile QC reports of sequencing statistics and files to aid in the immunogenomics review of candidates. The RunningPipeline_README (**Supplementary File 3**) document includes a detailed walkthrough to run the pipeline and download results back to your local machine.

### Hardware

ImmunoNX is designed to be flexible and can be launched from any Unix-based (Linux/Mac) system with a reliable high-speed internet connection. The pipeline execution is performed on Google Cloud Platform, where specific resource minimum quotas are recommended for optimal performance. These minimums include at least 100 CPUs and 100 preemptible CPUs, 25 in-use IP addresses, and 10 TB of Persistent Disk SSD storage. This cloud-based architecture ensures scalability and consistent performance regardless of local computing resources.

### Software

Several key software tools are required to run and interact with the pipeline. Version control for the repositories is managed with Git. Docker provides containerization to ensure portability and allow versioned image tags for version control of the underlying tools. Cromwell interprets the WDL scripts and orchestrates the allocation of compute resources and execution of the pipeline. The google-cloud-sdk is essential for interfacing with Google Cloud Platform services. For downstream immunogenomics review, R is required to run pVACview, while IGV enables visualization of genomic data. Microsoft Excel or similar spreadsheet software aids in viewing tabular output files.

# Procedure

The neoantigen prediction and prioritization protocol described here is broadly divided into the ImmunoNX computational pipeline, followed by the immunogenomics manual review.

## ImmunoNX pipeline step-by-step instructions

Before proceeding, the user must set up a Google Cloud account and complete user authentication. A brief walkthrough including steps for account setup and starting a pipeline test run with an example tumor cell line data is provided below. Detailed instructions are available in RunningPipeline_README.md (**Supplementary File 3**).

## Local setup

Create a directory where the pipeline set-up will be done and clone the git repositories that have the workflows, scripts to help run them, and template YAML files. These repositories are version controlled, and checking out a specific tagged release is recommended to ensure consistency across experiments.

For the example data here, we used the following branches for cloning the git repositories.

```
mkdir ~/hcc1395-immuno
cd hcc1395-immuno
mkdir git

git clone https://github.com/griffithlab/ImmunoNX_protocol.git
git clone --branch v1.4.0 https://github.com/wustl-oncology/analysis-wdls.git git/
git clone --branch v1.4.8 https://github.com/wustl-oncology/cloud-workflows.git git/
```

Google cloud access must be authenticated using Google cloud account credentials, after which a Google billing project can be set. The user must have the required IAM Roles to create and run resources in the selected project. The user must have the "Editor" and "Security Admin" IAM roles. A service account user with appropriate IAM roles will be configured in a subsequent step.

```
docker run -it google/cloud-sdk:latest /bin/bash
gcloud auth login #authenticate cloud access
gcloud config set project [Google Project] #set appropriate billing project
```

## Set up cloud service account, firewall settings, and storage bucket

The 'resources.sh' script from the cloud-workflows repository is used to set up the service account, firewall, and storage bucket. A storage bucket is a container in Google Cloud that stores and organizes files. The --project parameter provided here must match the same billing project configured above. Successful execution of this script creates two new configuration files in the current directory: 'cromwell.conf' and 'workflow_options.json'. The storage bucket created in this step can either be reused for multiple samples or projects, or a unique bucket can be created for individual projects or samples depending on data management preferences.

```
cd ~/hcc1395-immuno/git/cloud-workflows/manual-workflows
bash resources.sh init-project --project [Google Project] --bucket [Google Bucket]
```

## Download input data files

The example raw data files are downloaded from a public repository and untarred into a raw data directory as follows:

```
mkdir -p ~/hcc1395-immuno/raw_data
cd ~/hcc1395-immuno/raw_data
wget https://genomedata.org/hcc1395/fastqs/all/Exome_Norm.tar
wget https://genomedata.org/hcc1395/fastqs/all/Exome_Tumor.tar
wget https://genomedata.org/hcc1395/fastqs/all/RNAseq_Tumor.tar
tar -xvf Exome_Norm.tar
```

```
tar -xvf Exome_Tumor.tar
tar -xvf RNAseq_Tumor.tar
```

These TAR files contain whole exome and RNAseq data for the HCC1395 breast cancer cell line, and whole exome data for the matched normal HCC1395 lymphoblastoid line in FASTQ format[41].

## Create a copy of reference and annotation files in the Google Bucket

A custom and updated reference can be created using the instructions in RunningPipeline_README (**Supplementary File 3**), alternatively, a set of reference files (human genome build GRCh38, Ensembl v105) can be copied from the following publicly hosted bucket.

```
gsutil -u [Google Project] ls gs://griffith-lab-workflow-inputs/
gsutil -u [Google Project] cp -r gs://griffith-lab-workflow-inputs/human_GRCh38_ens105 gs://[User Google Bucket Reference Folder Path]
```

## Obtain an example configuration (YAML) file on local system

The template YAML for this data is provided in the 'ImmunoNX_protocol' repository cloned above. The YAML file must be updated with the appropriate paths for the input data including FASTQ files and reference files obtained in the previous two steps. If the YAML is set up for a different sample's data, the clinical HLA alleles (if available) and RNA strand settings should be updated along with the relevant file paths.

```
mkdir ~/hcc1395-immuno/yamls
cp ~/hcc1395-immuno/git/immuno_gcp_wdl_manuscript/template.yaml ~/hcc1395-immuno/yamls/hcc1395_local.yaml
```

## Stage input data files to cloud bucket

Raw data files on the local system are uploaded to the cloud bucket using the 'cloudize-workflow.py' script. Additionally, a YAML validator python script is provided in the docker image used to stage input files. This script can be run on the final YAML file to perform basic checks such as ensuring all file paths exist, raw FASTQ files are paired, and there are no empty YAML input arguments. Once the YAML file has been checked, the 'cloudize-workflow.py' script is run to parse the YAML file, upload any local files to the Google Bucket, and generate a Cloud YAML file with the updated file paths. The Cloud YAML file is uploaded to the Google Bucket.

```
cd ~/hcc1395-immuno
docker run -v $PWD:$PWD -w $PWD -it mgibio/cloudize-workflow:latest /bin/bash

python3 /opt/scripts/validate_immuno_yaml.py yamls/hcc1395_local.yaml

python3 /opt/scripts/cloudize-workflow.py [Google Bucket] git/analysis-
```

```
wdls/definitions/immuno.wdl yamls/hcc1395_local.yaml --output=yamls/hcc1395_cloud.yaml

gsutil cp yamls/hcc1395_cloud.yaml [Google Bucket]
```

### Start a Google VM that will run Cromwell and orchestrate completion of the workflow

The 'start.sh' script from the cloud-workflows repository is used to set up the Google VM.

```
cd [~/hcc1395-immuno/git/cloud-workflows/manual-workflows]
bash start.sh [Google Instance Name] --server-account [Google Service Account] --project [Google Project] --boot-disk-size=250GB
```

### Log into the VM and copy the YAML file from the bucket

The following commands are used to log onto the VM and check its status. Once the Google Compute Engine and Cromwell service have initiated successfully, the Cloud YAML file previously uploaded to the Google Cloud Bucket is localized to the virtual machine.

```
gcloud compute ssh [Google Instance Name]
journalctl -u google-startup-scripts -f
#WAIT FOR ... Finished Google Compute Engine Startup Scripts. ...

journalctl -u cromwell -f
#WAIT FOR ... INFO - Cromwell 88 service started on 0:0:0:0:0:0:0:0:8000 …

gsutil cp [Path to Cloud YAML on Google Bucket] .
```

### Run the immuno workflow and monitor progress

While logged into the Google Cromwell VM instance, the workflow is launched using the 'submit_workflow' command provided in the 'helpers.sh' script, which is automatically uploaded during VM setup. This command submits the analysis pipeline, 'immuno.wdl', along with the input data and arguments contained in the Cloud YAML file to the Cromwell service, thus initiating the pipeline run. Upon submission, Cromwell returns a unique workflow ID, which can be used to monitor progress and retrieve outputs; it is recommended to record the workflow ID at this step. The execution of a workflow typically takes 2-3 days depending on dataset size and resource requirements. Users may monitor its progress by inspecting the log file that is generated in real-time using the 'journalctl' command. If the run is successful, the bottom of the cromwell log will state that the workflow "completed with status Succeeded", along with the workflow ID. If the run has an error, then its status will say "Failed". A later section describes suggestions to troubleshoot "Failed" runs.

```
source /shared/helpers.sh
export wdl_path=/shared/analysis-wdls/definitions
```

```
submit_workflow $wdl_path/immuno.wdl ~/hcc1395_cloud.yaml

journalctl -f -u cromwell
```

### Save workflow metadata and retrieve results to a local machine

Once the workflow has completed, the 'save_artifacts' command can be run on the VM to save the workflow metadata to the Google Bucket.

```
source /shared/helpers.sh
save_artifacts [Workflow ID] [Google Bucket Path to save workflow metadata]
```

Results can be retrieved from the cluster using the 'pull_outputs.py' script, made available through the 'mgibio/cloudize-workflow' docker image. The 'outputs.json' file created by the prior step lists the files that will be downloaded and is the input to the script.

```
docker run -it -v $PWD:$PWD -w $PWD mgibio/cloudize-workflow:latest /bin/bash

# Download immuno output files to VM and copy to Google Bucket
mkdir ~/hcc1395-immuno/final_results
cd ~/hcc1395-immuno/final_results
python3 /opt/scripts/pull_outputs.py --outputs-file=[Google Bucket Path to save workflow metadata]/[Workflow ID]/outputs.json --outputs-dir=~/hcc1395-immuno/final_results
```

The results are organized as shown in **Figure 2**. Select from outputs each of the major steps, such as CRAM and BAM files of the DNA and RNA alignment steps, germline and somatic variant calling VCFs, gene, transcript and variant RNA expression estimates, and the neoantigen candidate predictions, are all saved to aid in vaccine design and to ask exploratory questions. The predictions can be visualized using pVACview on a local machine using RStudio. The command below will launch the pVACview R/Shiny App which has information on the files that must be uploaded.

```
R -e "shiny::runApp('pVACseq/mhc_i/', port=3333)"
```

### Immunogenomics review

The immunogenomics review is conducted in two stages and involves systematic evaluation of neoantigen candidates using both computational tools and manual assessment to ensure the highest quality vaccine design. The first stage begins with an evaluation of the case's characteristics and cancer driver variants (if any). This is followed by a thorough QC assessment of the raw sequencing data, examining metrics such as read depth, duplication rates, sample relatedness, contamination levels, and RNA specific alignment characteristics (e.g. transcript end bias and strandedness). The clinical and in-silico predicted HLA alleles are also compared between the normal and tumor DNA samples to ensure that no sample mix-ups occurred. Files generated by the computational pipeline are loaded into pVACview to assess each neoantigen candidate against predefined criteria, including DNA variant allele frequency (VAF), transcript support, binding affinity and presentation, agretopicity and RNA expression levels. Candidates arising from cancer driver genes are given special attention, with review criteria being slightly relaxed for these candidates, at the reviewers' discretion. Multi-algorithm support for binding predictions is evaluated, with additional scrutiny given to cases where algorithms disagree. Dinucleotide and complex variants may be flagged for additional genomic review in the second stage even if they have low RNA VAF/allele expression, as their complex structure may make automated read counting inaccurate. Candidates receive an 'Accept', 'Review', or 'Reject' designation along with optional comments via the pVACview interface, and these immunogenomics review decisions are exported in TSV and Excel formats to facilitate the second stage of review. A preliminary Genomics Review Report summarizing the QC findings and case characteristics is generated at this stage. In the second stage, each candidate is examined in IGV, where reviewers assess the somatic variant quality, check for proximal variants that may affect the predicted peptide sequence, verify RNA expression of the variant allele, and evaluate transcript isoform structure and expression. Our group has previously published guidelines[75] on reviewing somatic variants in IGV and a summary of the review criteria is provided in the ManualReview_README (**Supplementary File 4**).

The review culminates in the generation of a finalized Genomics Review Report that integrates findings from both stages, including annotated peptide sequences, manufacturing specifications, and an executive report detailing all findings and special considerations for individual candidates (see **Supplementary File 1** for example report). The candidates that make it through the review phases, can be formatted and annotated into an order form using pVACtools' 'pvacseq create_peptide_ordering_form' command (available as part of the pvactools Python package version 6.0. and above). This form includes the 51-amino acid sequence with the class I and class II peptides and variants annotated, along with a column for the peptide's molecular weight. This spreadsheet can be sent to a peptide manufacturer. Alternatively, to generate a DNA vector-based vaccine, the outputs can be processed using pVACvector.

At our institute, the first stage of the immunogenomics review is conducted at an Immunogenomics Tumor Board (ITB) meeting that brings together a multidisciplinary team including surgeons, oncologists, genomicists, bioinformaticians, and immunologists. During the meeting, the group first reviews the Genomics Review Report and discusses the type of cancer, driver variants, and QC metrics. Subsequently, the group prioritizes candidates, and records comments using pVACview. The resulting files are exported and used in the second stage of the immunogenomics review. Additional information about the review process and generating the peptide and DNA vector order forms is also available in the ManualReview_README (**Supplementary File 4**).

## Troubleshooting

If a pipeline run is submitted and it fails, this is often due to a user error in the YAML file. While the YAML validator script covers multiple commonly observed errors, it can't anticipate all possible errors. Sample mix-ups can be identified by reviewing genotype concordance results to ensure that the normal DNA, tumor DNA, and tumor RNA are from the same patient. File corruption issues that may occur during data transfer can be detected by checking MD5 checksums. Additionally, if RNA expression estimates seem inaccurate, confirming the RNA strandedness results can help identify potential strand-setting issues that can be corrected in the YAML file.

If working with unusually large datasets, resource requests in the WDLs can be edited prior to execution. This will be done by modifying the appropriate WDL in the analysis-wdls directory on the head node VM prior to launching the workflow. Note that the workflows.zip file must be regenerated using the 'refresh_zip_deps' command in the 'helpers.sh' script to incorporate any changes in the subsequent workflow run. In cases where tumor purity is low, the 'varscan_min_var_freq', 'fp_min_var_freq', and 'filter_somatic_llr_threshold' parameters in the YAML can be reduced from the default values to increase sensitivity. However, this increases the risk of including false positive variants and should be applied cautiously.

If the run fails with messages in the Cromwell logs indicating quota limits, this suggests the workflow has exceeded the project's current resource quota. Increasing quotas (CPUs or storage) in the Google Cloud console may be necessary. For new users, ensure that the correct IAM roles are granted (e.g., Editor and Security Admin) so that they can run workflows and access required project resources.

By default, the pipeline has Cromwell's call-caching feature enabled. While a user is working on the same head node VM where Cromwell is running, they can adjust the parameters that need to be fixed and initiate a new run. Cromwell's call-caching feature will reuse results for completed tasks that are not affected by the adjustments, effectively allowing the user to short-cut the completed steps, and thus saving time and compute costs.

For further guidance, please refer to the supplementary troubleshooting documentation Troubleshooting_README (**Supplementary File 5**).

## Anticipated results

Here we describe our workflow for prediction and prioritization of neoantigens for clinical trials. The Google Cloud implementation of ImmunoNX allows it to be easily portable, and since it is version controlled, one can simultaneously have a stable pipeline version for one trial or set of experiments while using a more up to date version for a newer trial.

ImmunoNX results and Immunogenomics review of HCC1395 example dataset

We executed ImmunoNX on the HCC1395 example dataset in 2 days,13 hours at a cost of $47.03 using the Google Cloud platform. The output files in the 'final_results' folder are organized based on the major workflows in the immunogenomics pipeline, with the primary files needed for review of neoantigen candidates located in the pVACseq folder, specifically in 'mhc_i' and 'mhc_ii' (for neoantigens prioritized based on class I and class II MHC alleles respectively). The results for HCC1395 were subjected to a comprehensive immunogenomics review, summarized in **Figure 4A**. 322 of the 1264 somatic variants gave rise to neoantigens, as the remaining variants did not directly result in protein coding changes. In practice, the first phase of the immunogenomics review, conducted by the ITB, aims to prioritize 20-25 candidates for a 16-candidate synthetic long peptide vaccine design (20-40 for DNA vector-based designs). However, for illustrative purposes, all 322 candidates for the HCC1395 dataset

were subjected to immunogenomics review. With no limits on the vaccine design, 90 candidates were prioritized by ITB, out of which 78 candidates were accepted after the immunogenomics review. The primary causes for a candidate's rejection were poor binding predictions (118) or lack of RNA expression (73) (**Figure 4B**). While the three variant calling algorithms largely agree, using multiple variant callers enabled the identification of more neoantigen candidates than if any one variant caller was used in isolation (**Figure 4C**). For peptide presentation, passing candidates were predicted for all four of the HLA alleles in this sample. Some algorithms, such as NetMHC, SMM, and SMMPMBEC did not provide binding predictions for the HLA-B*82:02 allele, highlighting the advantage of incorporating multiple binding prediction algorithms (**Supplementary Figure 2A-B**). Additionally, while all eight binding prediction algorithms often agreed that a candidate was a strong binder (29/322), most of the passing candidates were predicted to be strong binders by a subset of the algorithms for both class I (**Figure 4D**, **Supplementary Figure 2C**) and class II allele predictions. A few of the interesting candidates flagged during pVACview review in ITB and immunogenomics review are highlighted in **Figure 4E-G**. The SORBS3 candidate may have been accepted based on the standard metrics such as IC50 binding, DNA VAF, and gene expression in the summary table (**Figure 4E**). However, the reference match analysis done as part of pVACseq and visualized in pVACview showed that the mutant peptide sequence happened to match significantly with the wildtype peptide sequence of the SLC45A3 gene (**Figure 4F**). This match decreases the likelihood that this candidate would be targeted by the immune system due to immune tolerance. Conversely, the low variant allele expression of the JUP candidate suggests that it would be a poor vaccine candidate. However, the immunogenomics review in IGV revealed a 41bp deletion that was incorrectly being called as a splice event by the STAR aligner, and thus, was annotated as not being expressed by the pipeline (**Figure 4G, Supplementary Figure 2D**). Manual assessment using IGV was essential for rescuing this candidate. Detailed curation and review for every candidate in the HCC1395 dataset is available in **Supplementary Table 3**. A summary of interesting candidates, including the 12 candidates that were rejected in the IGV review phase is included at the end of the Genomics Review Report. Since a neoantigen vaccine has a limited number of candidates, it is imperative that we exclude candidates that are unlikely to stimulate an immune response, illustrating the importance of the IGV review step.

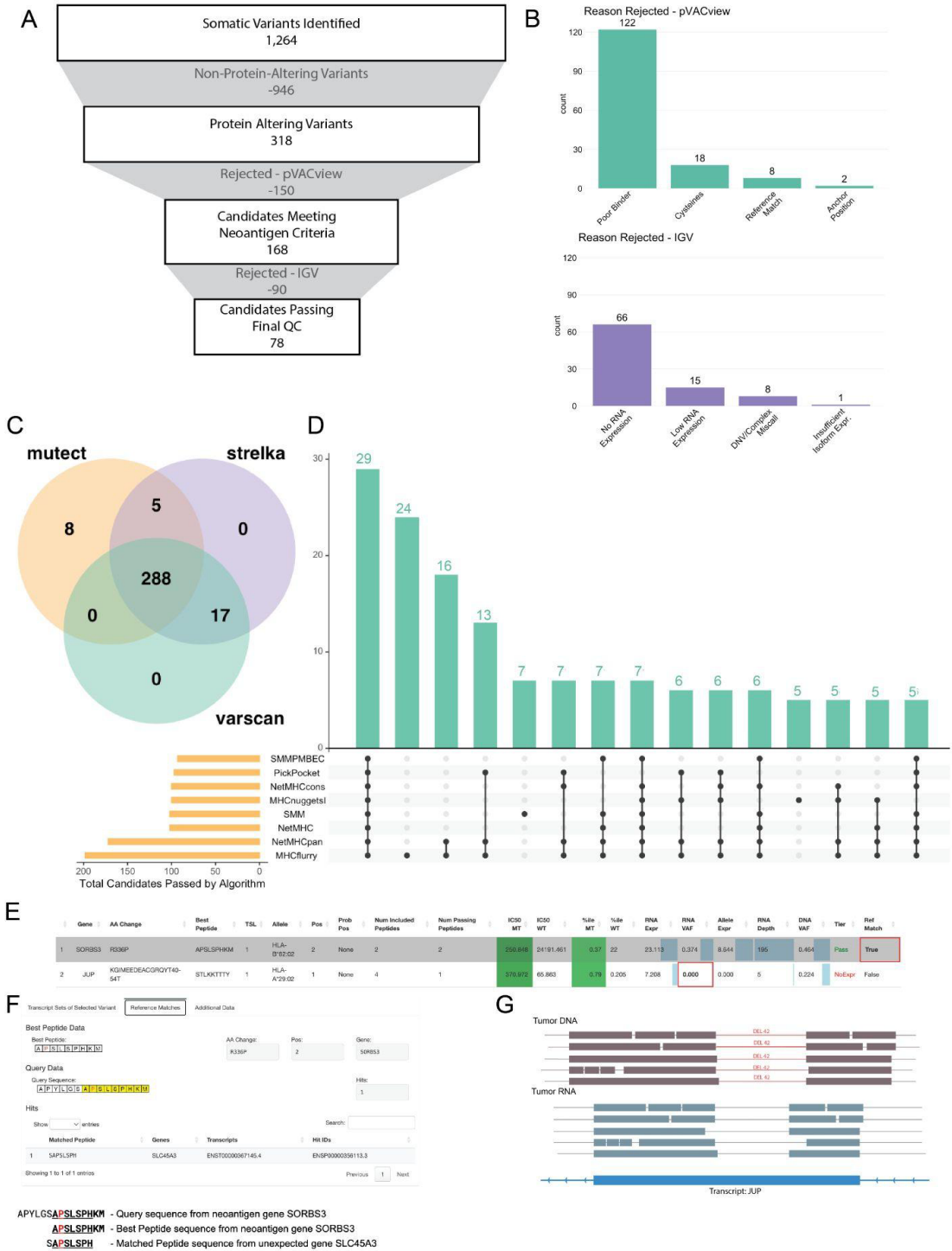

**Figure 4: Neoantigen prediction workflow results for HCC1395.** **(A)** Depiction of the main filtering steps from original somatic variants to final vaccine design. **(B)** Bar plots with additional details on why candidates were rejected based on epitope criteria (top) and variant criteria (bottom). **(C)** Venn diagram comparing agreement across three

variant calling algorithms. **(D)** Upset plot depicting class I binding prediction algorithms agreement for each peptide. **(E)** pVACview summary table with binding metrics, gene and variant expression and DNA VAF for SORBS3 and JUP neoantigen candidates. **(F)** pVACview reference match figure showing overlap of SORBS3 predicted neoantigen with wild-type SLC45A3 protein sequence. **(G)** Genome viewer-like depiction of the tumor DNA track (top), tumor RNA track (middle), and transcript track for part of the JUP gene. The 42bp deletion observed in the DNA is also observed in the RNA, however the STAR aligner annotates it as an intron, resulting in the incorrect RNA depth and RNA VAF values observed in **(E)**.

## Applications of the described workflow in clinical trials

We have used the workflow described here to design neoantigen cancer vaccines for over 185 patients, spanning multiple tumor types across 11 clinical trials (**Figure 5A**), suggesting that this workflow can be reliably used to identify targetable neoantigens agnostic of the tumor type. Cancers with higher tumor cellularity, such as breast cancer, were more likely to generate targetable neoantigens compared to tumor types such as pancreatic cancer where high tumor purity samples are more difficult to obtain (**Figure 5B**). Nonetheless, on average, patients in pancreatic cancer vaccine trials received a vaccine targeting 12 neoantigens.

Once the pipeline is initiated, it typically completes in 2 to 3 days, scaling largely based on the size of the FASTQ files and, to a lesser extent, Tumor Mutation Burden (**Figure 5C, 5D**). Overall, once the sequencing data is ready, the entire workflow is typically complete in 11-15 days, with the ITB and manual review taking 7-10 days (2-4 days with dedicated personnel) (**Figure 5D**). Therefore, the ImmunoNX pipeline and workflow presented here enable the design and ordering of a personalized neoantigen vaccine in less than three months after a patient is enrolled in a trial. While this timetable meets the needs of current vaccine trials, further optimizations could lead to shorter turnaround times.

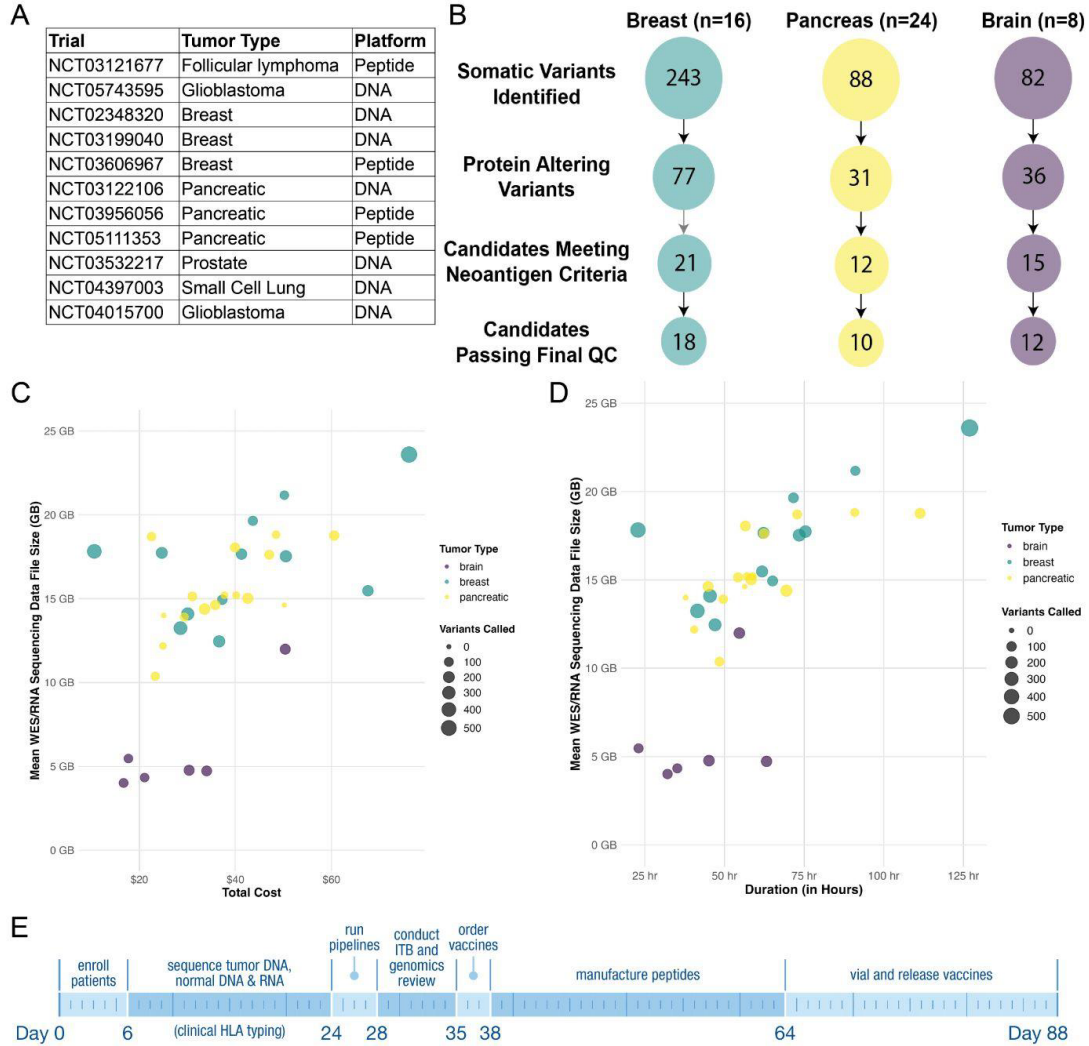

**Figure 5: Neoantigen prediction pipeline usage in clinical trials. (A)** Summary of clinical trials, including tumor types and vaccine platforms leveraging the neoantigen vaccine pipeline described here. **(B)** Cumulative summary of main filtering steps showing the average number of candidates per patient for each tumor type across a subset of patients with breast, pancreas, or brain cancer. **(C, D)** Scatterplots of mean WES DNA/RNA sequencing data file size for a patient by the total cost **(C)** and time **(D)** of running the computational pipeline. Point size scales with the number of variants called in the patient, and color is based on tumor type. **(E)** Exemplar clinical workflow timeline showing a typical complete timeline from patient consent and enrollment to vaccine vialing and release.

As personalized neoantigen vaccines become an increasingly sought-after and widely available immunotherapy, efficient and comprehensive procedures to design vaccines are essential. Here, we present a pipeline that takes tumor DNA/RNA and normal DNA sequencing data to produce high-quality neoantigen predictions. Our results demonstrate that a consensus-based approach, such as that used in HLA allele prediction, variant calling, and binding prediction, allows us to leverage the benefits and mitigate the weaknesses of multiple tools. Our results are organized for visual evaluation and interpretation with pVACview and IGV, and we have also outlined extensive criteria to guide immunogenomic review to ensure prioritization of the best candidates. We acknowledge that this manual review process can be time-consuming and requires significant coordination. However, we maintain that expert oversight is a fundamentally necessary step for creating safe and effective vaccines. As the field evolves, our pipeline is designed to be continually updated with new tools and adjusted to adapt to the future needs of the immuno-oncology community.

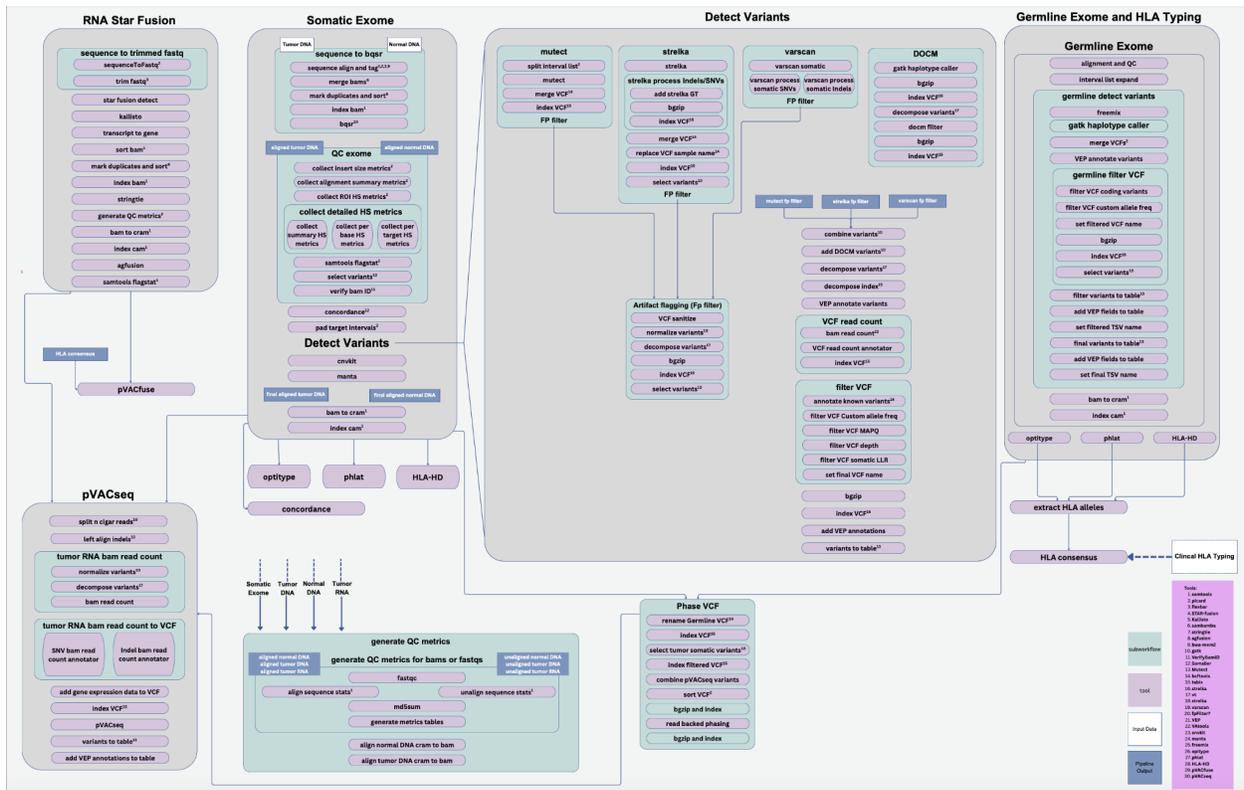

**Supplementary Figure 1: Tools used in the ImmunoNX pipeline.** The primary WDL that runs the ImmunoNX pipeline comprises multiple workflows such as 'Germline Exome and HLA Typing', 'Somatic Exome', etc. depicted in the large rectangles. Each workflow includes subworkflows depicted in green, that in turn run tasks or tools depicted in purple.

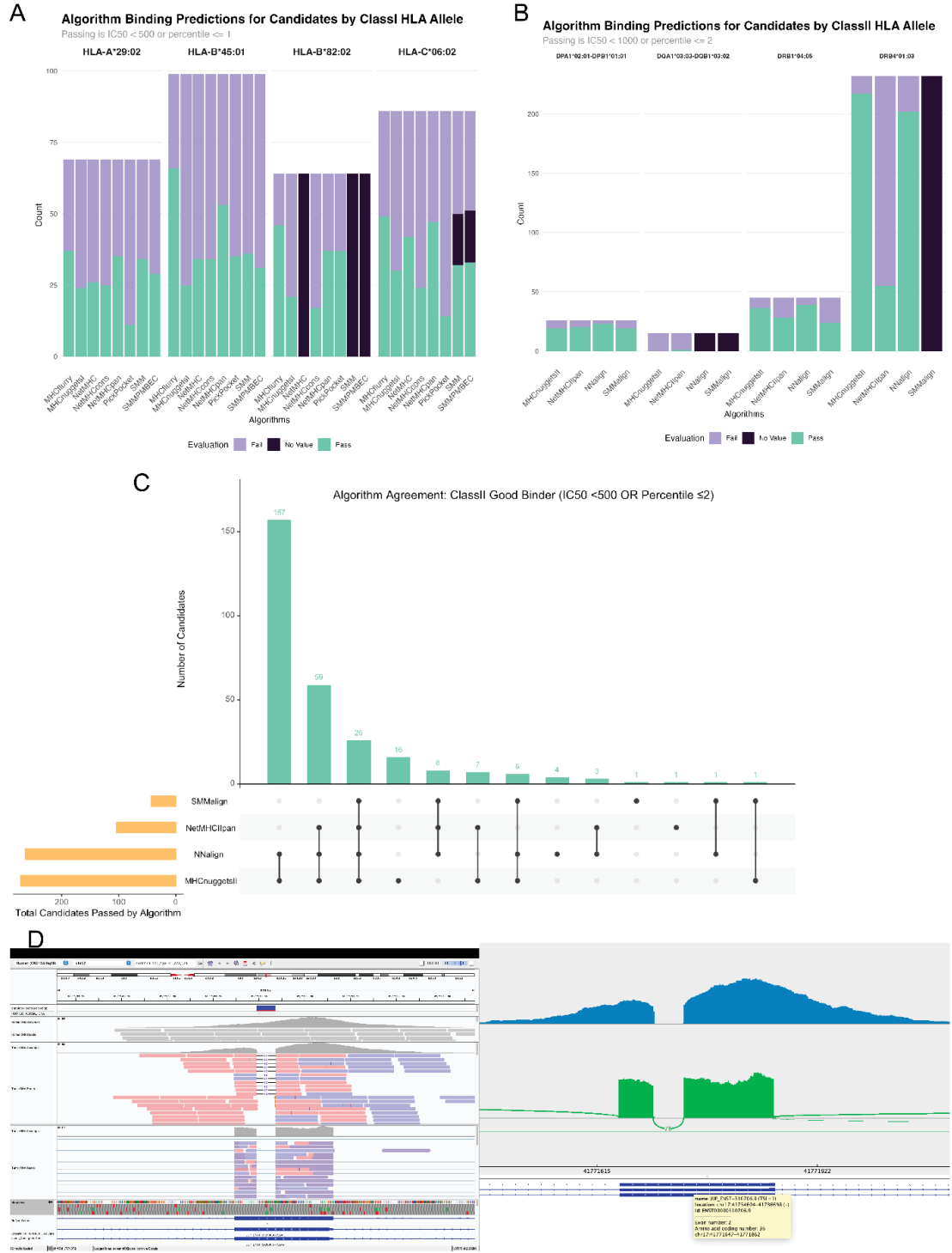

**Supplementary Figure 2: Neoantigen prediction workflow results for HCC1395**. **(A)** Class I and **(B)** Class II Bar charts grouped by HLA alleles showing each algorithm's binding predictions. Stacks within each bar show the number of candidates that were below the binding thresholds ('Pass' in green), above binding thresholds ('Fail' in lavender),

or did not give predictions ('No value' in black). **(C)** Upset plot depicting class II binding prediction algorithms agreement for each peptide. **(D)** IGV screenshot of the JUP candidate depicted in Figure 4H. On the left, the top track shows the normal DNA, middle track shows tumor DNA, and bottom track shows tumor RNA. On the right, the sashimi plot for visualizing transcript splice junctions is shown. Tumor DNA track shows 42bp deletion, while the tumor RNA track and sashimi plot suggest that the aligner considered the deletion an intron region and has spliced across it.

**Author contributions**
K.S. and E.S. wrote the manuscript and produced the visualizations. M.G. and O.L.G. conceived and developed the protocol. S.K., C.A.M, T.B.M., L.H., implemented and updated ImmunoNX. K.S., E.S., J.L., J.Y., M.H.H., M.K., I.R. analyzed and interpreted data. K.S., E.S., M.G., O.L.G., S.K., C.A.M., J.A.F.revised the manuscript. J.A.F., W.E.G., O.L.G., and M.G. participated in project administration and supervision, and provided resources. W.E.G. supervised clinical trials NCT02348320, NCT03199040, NCT03606967, NCT03122106, NCT03956056, NCT05111353, T.A.F. supervised clinical trials NCT03121677, T.M.J. supervised clinical trials NCT04015700 and NCT05743595, R.K.P. supervised clinical trials NCT03532217, J.P.W. supervised clinical trials NCT04397003.

**Acknowledgements**
We are grateful to the patients and their families for the donation of their samples and participation in clinical trials. M.G. was supported by the National Human Genome Research Institute (NHGRI) of the NIH under award number R00HG007940. M.G. and O.L.G. were supported by the NIH National Cancer Institute (NCI) under award numbers U01CA209936, U01CA231844, and U24CA237719. M.G. was supported by the V Foundation for Cancer Research under award number V2018-007. M.G., O.L.G., H.X., W.E.G., and T.A.F. were supported by the NCI under award number U01CA248235. T.A.F and W.E.G were supported by the Siteman Cancer Center (P30CA91842). T.A.F was also supported by the Leukemia SPORE (P50CA171063), Lymphoma Research Foundation, Blood Cancer United, the Howard Steinberg Family Foundation, and the Paula and Rodger Riney Blood Cancer Initiative. M.G., O.L.G. and W.E.G. were supported by the NCI/Leidos Biomedical Research Subcontract 20X012F5P50. S.P.G. was supported by the Foundation for Barnes-Jewish Hospital. K.S. was supported by NIH/NIGMS T32-GM139774. JAF was supported by the NCI under K22CA282364 and the Cancer Research Foundation (CRF) Young Investigator Award. W.E.G was supported by CA/NCI NIH T32-CA009621, Susan G. Komen for the Cure (KG111025), B101/Centene Corporation P19-00559, Alvin J. Siteman Cancer Center Investment Program grant 4035, CA/NCI NIH R01-CA240983, MedImmune AstraZeneca 004135-946897, P50 CA196510, and P50 CA272213. This work was also supported in part by the Washington University Institute of Clinical and Translational Sciences from the National Center for Advancing Translational Sciences (NCATS) of NIH under award number UL1TR002345. Finally, this work was supported by a gift from the Goldberg Family Foundation.

**Conflicts of interest**
K.S., E.S., K.C.C, M.G., and O.L.G. are consultants for the Jaime Leandro Foundation and Pathfinder Oncology. TAF holds equity in Wugen, Orca Bio, and Indapta Therapeutics; and has research funding from ImmunityBio, AI Proteins,

**Data availability**

All relevant data are within the paper and supplementary data. Raw sequence data for HCC1395 is available at https://genomedata.org/hcc1395/fastqs/all/. The supplementary data files and protocols referenced in this manuscript are uploaded to Zenodo and the following GitHub repository, https://github.com/griffithlab/ImmunoNX_protocol. Users may submit questions or requests related to the vaccine design workflow on this GitHub repository.

**Code availability**

All WDL files for the computational pipeline are in the GitHub repository, https://github.com/wustl-oncology/analysis-wdls (MIT license). The helper scripts and files used for running the pipeline on the Google Cloud are in the GitHub repository, https://github.com/wustl-oncology/cloud-workflows.

**References**

1. Schumacher, T. N. & Schreiber, R. D. Neoantigens in cancer immunotherapy. *Science* **348**, 69–74 (2015).

2. Weber, J. S. *et al.* Individualised neoantigen therapy mRNA-4157 (V940) plus pembrolizumab versus pembrolizumab monotherapy in resected melanoma (KEYNOTE-942): a randomised, phase 2b study. *Lancet* **403**, 632–644 (2024).

3. Hundal, J. *et al.* pVAC-Seq: A genome-guided in silico approach to identifying tumor neoantigens. *Genome Med.* **8**, 11 (2016).

4. Hundal, J. *et al.* Accounting for proximal variants improves neoantigen prediction. *Nat. Genet.* **51**, 175–179 (2019).

5. Hundal, J. *et al.* pVACtools: A Computational Toolkit to Identify and Visualize Cancer Neoantigens. *Cancer Immunol Res* **8**, 409–420 (2020).

6. Xia, H. *et al.* Computational prediction of MHC anchor locations guides neoantigen identification and prioritization. *Sci Immunol* **8**, eabg2200 (2023).

7. Xia, H. *et al.* pVACview: an interactive visualization tool for efficient neoantigen prioritization and selection. *Genome Med.* **16**, 132 (2024).

8. Richters, M. M. *et al.* Best practices for bioinformatic characterization of neoantigens for clinical utility. *Genome Med.* **11**, 56 (2019).

9. De Mattos-Arruda, L. *et al.* Neoantigen prediction and computational perspectives towards clinical benefit: recommendations from the ESMO Precision Medicine Working Group. *Ann. Oncol.* **31**, 978–990 (2020).

10. Ramirez, C. A. *et al.* Neoantigen Landscape Supports Feasibility of Personalized Cancer Vaccine for Follicular Lymphoma. *Blood Adv* (2024) doi:10.1182/bloodadvances.2022007792.

11. Zhang, X. *et al.* Neoantigen DNA vaccines are safe, feasible, and induce neoantigen-

specific immune responses in triple-negative breast cancer patients. *Genome Med.* **16**, 131 (2024).

12. Li, L. *et al.* Optimized polyepitope neoantigen DNA vaccines elicit neoantigen-specific immune responses in preclinical models and in clinical translation. *Genome Med.* **13**, 56 (2021).

13. King, D. A. *et al.* Complete Remission of Widely Metastatic Human Epidermal Growth Factor Receptor 2-Amplified Pancreatic Adenocarcinoma After Precision Immune and Targeted Therapy With Description of Sequencing and Organoid Correlates. *JCO Precis Oncol* **7**, e2100489 (2023).

14. Carreno, B. M. *et al.* Cancer immunotherapy. A dendritic cell vaccine increases the breadth and diversity of melanoma neoantigen-specific T cells. *Science* **348**, 803–808 (2015).

15. Cai, Z. *et al.* Personalized neoantigen vaccine prevents postoperative recurrence in hepatocellular carcinoma patients with vascular invasion. *Mol. Cancer* **20**, 164 (2021).

16. Freed-Pastor, W. A. *et al.* The CD155/TIGIT axis promotes and maintains immune evasion in neoantigen-expressing pancreatic cancer. *Cancer Cell* **39**, 1342–1360.e14 (2021).

17. Westcott, P. M. K. *et al.* Low neoantigen expression and poor T-cell priming underlie early immune escape in colorectal cancer. *Nat. Cancer* **2**, 1071–1085 (2021).

18. Zhang, M. *et al.* Clonal architecture in mesothelioma is prognostic and shapes the tumour microenvironment. *Nat. Commun.* **12**, 1751 (2021).

19. Garsed, D. W. *et al.* The genomic and immune landscape of long-term survivors of high-grade serous ovarian cancer. *Nat. Genet.* **54**, 1853–1864 (2022).

20. Newell, F. *et al.* Multiomic profiling of checkpoint inhibitor-treated melanoma: Identifying predictors of response and resistance, and markers of biological discordance. *Cancer Cell* **40**, 88–102.e7 (2022).

21. Liu, T. *et al.* High-affinity neoantigens correlate with better prognosis and trigger potent antihepatocellular carcinoma (HCC) activity by activating CD39+CD8+ T cells. *Gut* **70**,

1965–1977 (2021).

22. Thibaudin, M. *et al.* First-line durvalumab and tremelimumab with chemotherapy in RAS-mutated metastatic colorectal cancer: a phase 1b/2 trial. *Nat. Med.* **29**, 2087–2098 (2023).

23. Kinnersley, B. *et al.* Genomic landscape of diffuse glioma revealed by whole genome sequencing. *Nat. Commun.* **16**, 4233 (2025).

24. Yang, C. *et al.* Identification of tumor rejection antigens and the immunologic landscape of medulloblastoma. *Genome Med.* **16**, 102 (2024).

25. Bateman, N. W. *et al.* Proteogenomic analysis of enriched HGSOC tumor epithelium identifies prognostic signatures and therapeutic vulnerabilities. *NPJ Precis. Oncol.* **8**, 68 (2024).

26. Meghani, K. *et al.* Genomic and transcriptomic profiling of high-risk bladder cancer reveals diverse molecular and microenvironment ecosystems. *bioRxiv* 2024.12.21.629010 (2024) doi:10.1101/2024.12.21.629010.

27. Lin, M. *et al.* Neoantigen landscape in metastatic nasopharyngeal carcinoma. *Theranostics* **11**, 6427–6444 (2021).

28. Calsina, B. *et al.* Genomic and immune landscape Of metastatic pheochromocytoma and paraganglioma. *Nat. Commun.* **14**, 1122 (2023).

29. Wang, G. *et al.* CRISPR-GEMM pooled mutagenic screening identifies KMT2D as a major modulator of immune checkpoint blockade. *Cancer Discov.* **10**, 1912–1933 (2020).

30. Zhao, J. *et al.* Immune and genomic correlates of response to anti-PD-1 immunotherapy in glioblastoma. *Nat. Med.* **25**, 462–469 (2019).

31. Hollern, D. P. *et al.* B cells and T follicular helper cells mediate response to checkpoint inhibitors in high mutation burden mouse models of breast cancer. *Cell* **179**, 1191–1206.e21 (2019).

32. Trivedi, V. *et al.* mRNA-based precision targeting of neoantigens and tumor-associated antigens in malignant brain tumors. *Genome Med.* **16**, 17 (2024).

33. Gschwind, A. & Ossowski, S. AI model for predicting anti-PD1 response in melanoma using multi-omics biomarkers. *Cancers (Basel)* **17**, 714 (2025).

34. Bigelow, E. *et al.* A random forest genomic classifier for tumor agnostic prediction of response to anti-PD1 immunotherapy. *Cancer Inform.* **21**, 11769351221136081 (2022).

35. Borden, E. S. *et al.* NeoScore integrates characteristics of the neoantigen:MHC class I interaction and expression to accurately prioritize immunogenic neoantigens. *J. Immunol.* **208**, 1813–1827 (2022).

36. Rieder, D. *et al.* nextNEOpi: a comprehensive pipeline for computational neoantigen prediction. *Bioinformatics* **38**, 1131–1132 (2022).

37. Zou, B. *et al.* Integrative genomic analyses of 1,145 patient samples reveal new biomarkers in esophageal squamous cell carcinoma. *Front. Mol. Biosci.* **8**, 792779 (2021).

38. Nicholas, B. *et al.* Identification of neoantigens in oesophageal adenocarcinoma. *Immunology* **168**, 420–431 (2023).

39. Hashimoto, S. *et al.* Neoantigen prediction in human breast cancer using RNA sequencing data. *Cancer Sci.* **112**, 465–475 (2021).

40. Álvarez-Prado, Á. F. *et al.* Immunogenomic analysis of human brain metastases reveals diverse immune landscapes across genetically distinct tumors. *Cell Rep. Med.* **4**, 100900 (2023).

41. Griffith, M. *et al.* Genome Modeling System: A Knowledge Management Platform for Genomics. *PLoS Comput. Biol.* **11**, e1004274 (2015).

42. Voss, K., Auwera, G. & Gentry, J. Full-stack genomics pipelining with GATK4 + WDL + Cromwell. *F1000Research* **6**, (2017).

43. Sherry, S. T. *et al.* dbSNP: the NCBI database of genetic variation. *Nucleic Acids Res.* **29**, 308–311 (2001).

44. Landrum, M. J. *et al.* ClinVar: public archive of relationships among sequence variation and human phenotype. *Nucleic Acids Res.* **42**, D980–5 (2014).

45. Karczewski, K. J. *et al.* The mutational constraint spectrum quantified from variation in 141,456 humans. *Nature* **581**, 434–443 (2020).

46. Robinson, J. T. *et al.* Integrative genomics viewer. *Nat. Biotechnol.* **29**, 24–26 (2011).

47. Cibulskis, K. *et al.* Sensitive detection of somatic point mutations in impure and heterogeneous cancer samples. *Nat. Biotechnol.* **31**, 213–219 (2013).

48. Kim, S. *et al.* Strelka2: fast and accurate calling of germline and somatic variants. *Nat. Methods* **15**, 591–594 (2018).

49. Koboldt, D. C. *et al.* VarScan 2: somatic mutation and copy number alteration discovery in cancer by exome sequencing. *Genome Res.* **22**, 568–576 (2012).

50. Ainscough, B. J. *et al.* DoCM: a database of curated mutations in cancer. *Nat. Methods* **13**, 806–807 (2016).

51. Poplin, R. *et al.* Scaling accurate genetic variant discovery to tens of thousands of samples. *bioRxiv* 201178 (2017) doi:10.1101/201178.

52. Szolek, A. *et al.* OptiType: precision HLA typing from next-generation sequencing data. *Bioinformatics* **30**, 3310–3316 (2014).

53. Kumar, D. *et al.* Parental understanding of infant health information: health literacy, numeracy, and the Parental Health Literacy Activities Test (PHLAT). *Acad. Pediatr.* **10**, 309–316 (2010).

54. Kawaguchi, S., Higasa, K., Shimizu, M., Yamada, R. & Matsuda, F. HLA-HD: An accurate HLA typing algorithm for next-generation sequencing data. *Hum. Mutat.* **38**, 788–797 (2017).

55. Bray, N. L., Pimentel, H., Melsted, P. & Pachter, L. Near-optimal probabilistic RNA-seq quantification. *Nat. Biotechnol.* **34**, 525–527 (2016).

56. Pertea, M. *et al.* StringTie enables improved reconstruction of a transcriptome from RNA-seq reads. *Nat. Biotechnol.* **33**, 290–295 (2015).

57. Haas, B. J. *et al.* STAR-fusion: Fast and accurate fusion transcript detection from RNA-

Seq. *bioRxiv* (2017) doi:10.1101/120295.

58. Murphy, C. & Elemento, O. AGFusion: annotate and visualize gene fusions. *bioRxiv* 080903 (2016) doi:10.1101/080903.

59. Tang, Y. *et al.* TruNeo: an integrated pipeline improves personalized true tumor neoantigen identification. *BMC Bioinformatics* **21**, 532 (2020).

60. Zhou, C. *et al.* pTuneos: prioritizing tumor neoantigens from next-generation sequencing data. *Genome Med.* **11**, 67 (2019).

61. Huber, F. *et al.* A comprehensive proteogenomic pipeline for neoantigen discovery to advance personalized cancer immunotherapy. *Nat. Biotechnol.* 1–13 (2024).

62. Bais, P., Namburi, S., Gatti, D. M., Zhang, X. & Chuang, J. H. CloudNeo: a cloud pipeline for identifying patient-specific tumor neoantigens. *Bioinformatics* **33**, 3110–3112 (2017).

63. Vensko, S. P. *et al.* LENS: Landscape of Effective Neoantigens Software. *Bioinformatics* **39**, (2023).

64. Schenck, R. O., Lakatos, E., Gatenbee, C., Graham, T. A. & Anderson, A. R. A. NeoPredPipe: high-throughput neoantigen prediction and recognition potential pipeline. *BMC Bioinformatics* **20**, 264 (2019).

65. Kodysh, J. & Rubinsteyn, A. OpenVax: An Open-Source Computational Pipeline for Cancer Neoantigen Prediction. in *Bioinformatics for Cancer Immunotherapy: Methods and Protocols* (ed. Boegel, S.) 147–160 (Springer US, New York, NY, 2020).

66. Shi, Y., Jing, B. & Xi, R. Comprehensive analysis of neoantigens derived from structural variation across whole genomes from 2528 tumors. *Genome Biol.* **24**, 169 (2023).

67. Shah, N. M. *et al.* Pan-cancer analysis identifies tumor-specific antigens derived from transposable elements. *Nat. Genet.* **55**, 631–639 (2023).

68. Vita, R. *et al.* The Immune Epitope Database (IEDB): 2018 update. *Nucleic Acids Res.* **47**, D339–D343 (2019).

69. Fuchs, K. J. *et al.* Ribosome profiling shows variable sensitivity to detect open reading

frames for conventional and different types of cryptic T cell antigens. *Mol. Ther. Methods Clin. Dev.* **33**, 101391 (2025).

70. Cai, Y. *et al.* Immunopeptidomics-guided discovery and characterization of neoantigens for personalized cancer immunotherapy. *Sci. Adv.* **11**, eadv6445 (2025).

71. Kacen, A. *et al.* Post-translational modifications reshape the antigenic landscape of the MHC I immunopeptidome in tumors. *Nat. Biotechnol.* **41**, 239–251 (2023).

72. 1000 Genomes Project Consortium *et al.* A global reference for human genetic variation. *Nature* **526**, 68–74 (2015).

73. Miller, C. A. *et al.* Failure to detect mutations in U2AF1 due to changes in the GRCh38 reference sequence. *J. Mol. Diagn.* **24**, 219–223 (2022).

74. Dyer, S. C. *et al.* Ensembl 2025. *Nucleic Acids Res.* **53**, D948–D957 (2025).

75. Barnell, E. K. *et al.* Standard operating procedure for somatic variant refinement of sequencing data with paired tumor and normal samples. *Genet. Med.* **21**, 972–981 (2019).